\documentclass[11pt]{article}

\usepackage{cite}
\usepackage{amsmath}
\usepackage{amsfonts}
\usepackage{amssymb}
\usepackage{epsf}

\def\bbox#1{\mbox{\boldmath$#1$}}

\begin{document}

\bibliographystyle{myprsty}

\begin{center}
\begin{tabular}{c}
\hline
\rule[-5mm]{0mm}{15mm}
{\Large \sf A Problematic Set of}\\
\rule[-5mm]{0mm}{15mm}
{\Large \sf Two--Loop Self--Energy Corrections}\\
\hline
\end{tabular}
\end{center}

\vspace{1cm}

\begin{center}
Ulrich~D.~Jentschura, J\"{o}rg Evers and Christoph~H.~Keitel
\end{center}

\begin{center}
{\it Fakult\"{a}t Physik der Albert--Ludwigs--Universit\"{a}t,
Theoretische Quantendynamik,}\\
{\it Hermann--Herder--Stra\ss{}e 3, D--79104 Freiburg, Germany}\\[1ex]
\end{center}

\vspace{0.2cm}

\begin{center}
Krzysztof~Pachucki
\end{center}

\begin{center}
{\it Institute of Theoretical Physics,
University of Warsaw,}\\
{\it ul.~Ho\.{z}a 69, 00-681 Warsaw, Poland}\\[2ex]
\end{center}

\vspace{1.3cm}

\begin{center}
\begin{minipage}{10.5cm}
{\underline{Abstract}} We investigate a specific set of
two-loop self-energy corrections involving squared decay rates
and point out that their interpretation 
is highly problematic. The corrections cannot be interpreted
as radiative energy shifts in the usual sense. Some of the 
problematic corrections find a natural interpretation
as radiative nonresonant corrections to the natural line shape. They
cannot uniquely be associated with one and only one atomic level.
While the problematic corrections are rather tiny when 
expressed in units of frequency (a few Hertz for hydrogenic P
levels) and do not affect the reliability of quantum electrodynamics
at the current level of experimental accuracy, they may be of
importance for future experiments. The problems are connected with 
the limitations of the so-called asymptotic-state approximation
which means that atomic in- and out-states in the $S$-matrix
are assumed to have an infinite lifetime.
\end{minipage}
\end{center}

\vspace{1.3cm}

\noindent
{\underline{PACS numbers}}
12.20.Ds, 31.15.-p, 31.30Jv.\\
{\underline{Keywords}} quantum electrodynamics - specific calculations;\\
calculations and mathematical techniques in atomic and molecular physics;\\
Relativistic and quantum electrodynamic effects in atoms and molecules.

\newpage
%
% Introduction
%
\section{Introduction}
\label{Introduction}

In view of the 
rapid progress in ultra-accurate Lamb shift measurements in 
atomic hydrogen~\cite{BeEtAl1997,NiEtAl2000}, 
it appears useful to investigate the mathematical 
foundations of the theorems on which our level-shift calculations
are based, in addition to the continuing 
efforts (e.g.~\cite{Pa1994prl,JeMoSo1999,MeRi2000,Ye2001,Pa2001}) of evaluating
higher-and-higher order radiative corrections to the bound-state
energy levels. Here, we focus on a number of mathematical
subtleties connected with the famous theorem of 
Gell--Mann, Low and Sucher~\cite{GMLo1951,Su1957}
that forms the basis for the derivation of the expressions
investigated in level-shift calculations. This theorem
has historically proven to be an extremely useful tool in the 
analysis of bound-state quantum electrodynamics, and it yields
formal expressions for the renormalized level shifts
whose numerical and analytic evaluation has lead to 
accurate predictions for the bound-state energy levels
that are in agreement with all experiments reported so far
in the literature. However, second thoughts about some mathematical
subtleties connected with the theorem may be required
at the level of accuracy envisaged in projected experiments.

Consider the formula for the energy shift of an atomic state
as given by the well-known theorem of 
Gell--Mann, Low and Sucher~\cite{GMLo1951,Su1957}
\begin{equation}
\label{GMLow}
\Delta E_n = \lim_{\begin{array}{l}
{\scriptstyle \varepsilon \to 0^+}\\[-0.7ex] 
{\scriptstyle \lambda \to 1}\end{array}} \,
\frac{{\mathrm i} \, \varepsilon \, \lambda}{2} \,
\frac{\partial}{\partial \lambda} 
\ln\left[  \langle n | S_{\varepsilon,\lambda} | n \rangle_c \right]\,,
\end{equation}
where $|n\rangle$ is an unperturbed asymptotic bound-electron state
as given by the Dirac theory. $S_{\varepsilon,\lambda}$ is the
infinitesimally damped $S$ matrix given by the time-ordered
exponential
\begin{equation}
\label{Smatrix}
S_{\varepsilon,\lambda} = 
{\mathrm T} \exp\left( -{\mathrm i} \lambda \, \int_{-\infty}^\infty 
{\mathrm d} t \int {\mathrm d}^3 x \,
\exp(-\varepsilon |t|) \, {\mathcal H}_{\mathrm I}(x)
\right)
\end{equation}
where $x$ denotes the four-vector $(t, \bbox{x})$, 
and the interaction Hamiltonian density reads
\begin{equation}
\label{Hint}
{\mathcal H}_{\mathrm I}(\bbox{x}) =
- \frac{e}{2} \, \hat{A}_\mu(x) \, [\hat{\bar{\psi}}(x), \gamma^\mu 
\hat{\psi}(x)] - \frac12 \, \delta m \, [\hat{\bar{\psi}}(x), \hat{\psi}(x)]\,.
\end{equation}
The interaction Hamiltonian density
involves the quantized electromagnetic field
$\hat{A}_\mu(x)$ , the quantized Dirac
field $\hat{\psi}(x)$ in the Furry picture, 
and the mass counter term $\delta m$. The index $c$ in (\ref{GMLow})
indicates that only connected graphs~(see e.g. Ch.~6 of~\cite{ItZu1980})
enter into the expression for the energy shift.
Note that in writing down the 
expression (\ref{GMLow}), we implicitly assume the physical existence
of the asymptotic state $|n\rangle$, i.e.~of the unperturbed state
$|n\rangle$ with an infinite lifetime. 
If the interaction with the quantized electromagnetic field 
(the ``vacuum modes'') could be
``switched off'' [as it is assumed for the damped interaction 
(\ref{Smatrix}) in the distant past and future],
then all states would be asymptotic states and could be used to 
construct $S$-matrix elements rigorously. 
While this is -- strictly speaking --
unphysical, Eq.~(\ref{Smatrix}) is still an excellent
approximation for most bound-state calculations. 

A second-order evaluation of (\ref{GMLow}) in powers of
${\mathcal H}_{\mathrm I}$, which involves
the one-loop self-energy, shows that the radiative energy shift of an
excited state with nonvanishing angular momentum (say, a P state)
has an imaginary part
generated by the interaction of the atomic state with the quantized
electromagnetic field (which gives the expression for the 
decay rate $\Gamma$). This effect limits the validity 
of the asymptotic-state approximation which is a good approximation
as long as $\Gamma$ is much smaller than the separation
of atomic energy levels [see also the discussion
in Ch.~XXI of~\cite{Bo1979}, especially Eq.~(3.28) on p.~547 
{\em ibid.}]. 

Certain two-loop self-energy contributions to the 
hydrogenic energy levels involve
the {\em square} of the imaginary parts of two one-loop insertions,
which result in a shift of the real energy eigenvalue.
The error initially made in ignoring the decay width of the resonances
thus influences the real part of the eigenvalues which theory
predicts at two-loop order. This leads to a problem in connection
with the asymptotic-state approximation
that is originally used in writing down the expression (\ref{GMLow}).
At present, we have no better 
way to gauge the magnitude of this problematic effect  
but to evaluate it, according the current formalism, within a theory 
that {\em a priori} involves asymptotic states.
The effects discussed here touch a certain question
regarding the mathematical foundations of bound-state
quantum electrodynamics. 

This paper is organized as follows: in Sec.~\ref{2Pstate},
we present an evaluation of the problematic two-loop corrections
involving the squared decay rates for the 2P state of atomic 
hydrogen. In Sec.~\ref{Interpretation}, we argue that some of
the problematic corrections find a natural interpretation as
radiative corrections to the off-resonance effects that 
influence the line shape in atomic transitions.
Conclusions are left to Sec.~\ref{Conclusions}.

%
% A Concrete Example
%
\section{A Concrete Example}
\label{2Pstate}

\begin{figure}[htb]
\begin{center}
\begin{minipage}{12cm}
\centerline{\mbox{\epsfysize=7.7cm\epsffile{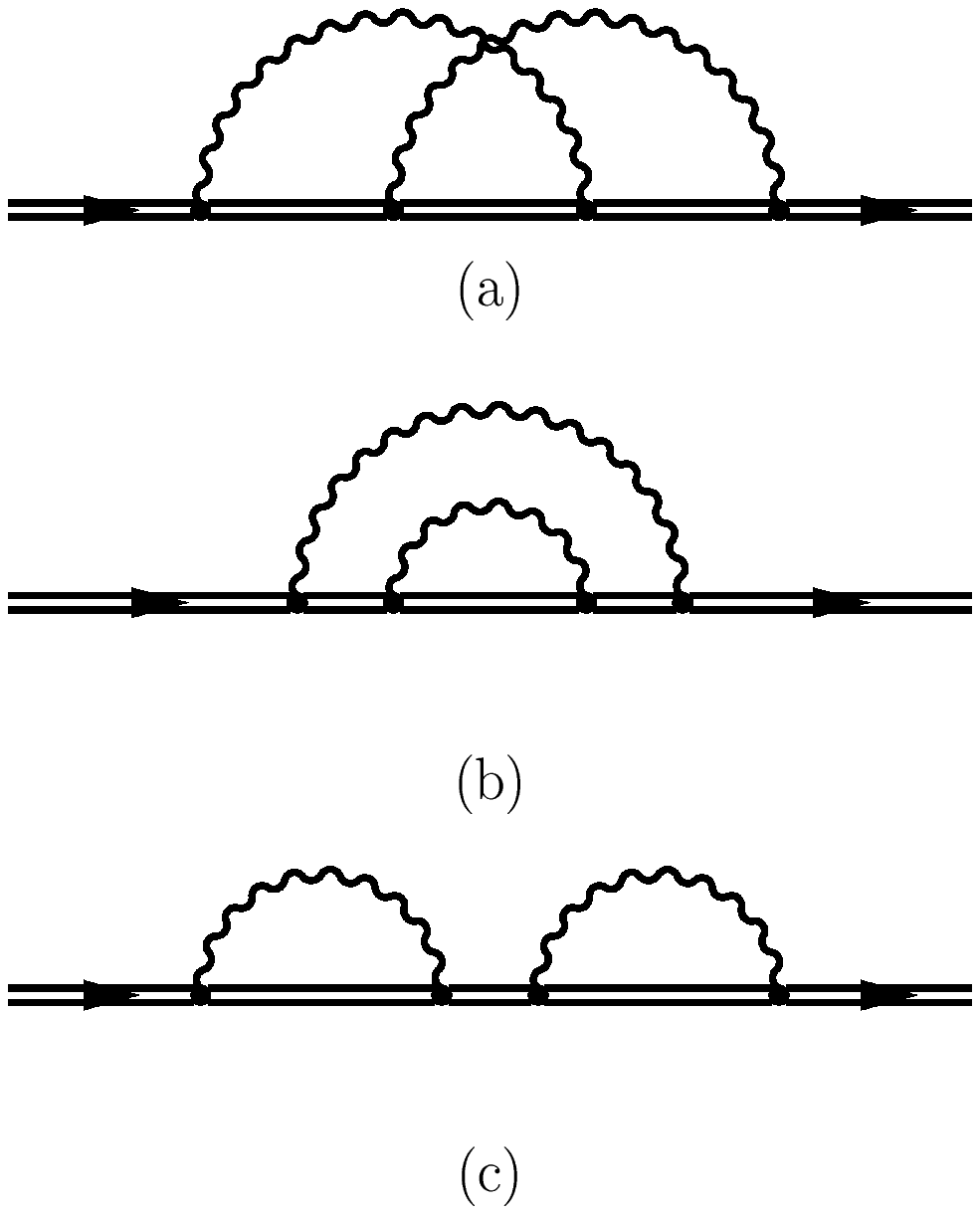}}}
\caption{\label{fig1} The crossed-loop (a), the rainbow diagram
(b) with one loop inside the other, and the
loop-after-loop diagram (c) which contribute to the two-loop
self-energy for a bound electron. The propagator of the bound
electron is denoted by a double line, and the arrow of time is
from left to right.}
\end{minipage}
\end{center}
\end{figure}

An expression for the two-loop self-energy correction to 
the energy of a bound hydrogenic state within 
the formalism of nonrelativistic quantum electrodynamics (NRQED)
has been derived in~\cite{Pa2001} [see Eq.~(16) {\em ibid.}],
and we take this equation as the starting point of our
investigations. Diagrammatically, 
the two-loop self-energy can be represented as in Fig.~\ref{fig1}.
We will investigate the effect within the $\epsilon$-method
that involves a scale-separation parameter
$\epsilon$ for the photon energy~(this method is explained in~\cite{Pa1993}
and the Appendix of~\cite{JePa2002}).
For the two photons with energies $\omega_1$ and $\omega_2$,
we need two scale-separation parameters $\epsilon_1$ and $\epsilon_2$.
These cancel when the high- and the low-energy
parts are added, and it is permissible to keep only the
divergent and constant terms as both $\epsilon_1 \to 0$ and
$\epsilon_2 \to 0$.
We therefore keep $\epsilon_{1/2}$ as variables and
evaluate only those contributions to the two-loop self-energy
integrals that correspond to the square of the
residues along the $\omega_1$ and $\omega_2$ integrations
(these correspond to the ``squared imaginary parts'' 
or ``squared decay rates''). We focus on the 2P state of atomic hydrogen
and use natural Gaussian (``microscopic'') units with 
$\hbar = c = \epsilon_0 = 1$. The Schr\"{o}dinger energy of an atomic
state is $E_n = - (Z\alpha)^2 \, m/(2 n^2)$
where $n$ is the principal quantum number,  
$\alpha$ is the fine-structure constant, and $Z$ is the nuclear
charge number.

%
% Fig. 1
%

There is a contribution due to the diagram with crossed loops
in Fig.~\ref{fig1} (a),
\begin{eqnarray}
\lefteqn{
\label{T1}
{\mathcal T}_1(2{\rm P})
= \lim_{\delta \to 0^+} - \left( \frac{2 \, \alpha}{3 \,\pi\,m^2} \right)^2 \,
\int_0^{\epsilon_1} {\rm d}\omega_1 \, \omega_1 \,
\int_0^{\epsilon_2} {\rm d}\omega_2 \, \omega_2 \,} \nonumber\\[2ex]
& & 
\times 
\left< 2{\rm P} \left|  p^i \, 
\frac{1}{H - {\rm i}\,\delta - E_{\rm 2P} + \omega_1} \, p^j \,
\frac{1}{H - E_{\rm 2P} + \omega_1 + \omega_2} \, p^i \,
\frac{1}{H - {\rm i}\,\delta - E_{\rm 2P} + \omega_2} \, 
p^j \right| 2{\rm P} \right>\,.
\end{eqnarray}
Here, the $p^i$ are the momentum
operators, and $H$ is the Schr\"{o}dinger Hamiltonian. 
We ``pick up'' only the terms of the ``squared-decay'' type, i.e.~the terms
generated by the infinitesimal half-circles around the poles
at $\omega_1 = E_{\rm 2P} - E_{\rm 1S}$ and 
$\omega_2 = E_{\rm 2P} - E_{\rm 1S}$. For the evaluation
of these terms, the specification of the infinitesimal imaginary
part $- {\rm i}\,\delta$ is required in order to fix the sign 
of the pole contribution. For the contribution $C_1(2{\rm P})$ 
generated by the poles at  $\omega_1 = E_{\rm 2P} - E_{\rm 1S}$ and 
$\omega_2 = E_{\rm 2P} - E_{\rm 1S}$ in ${\mathcal T}_1(2{\rm P})$, we obtain
\begin{eqnarray}
\label{C1}
C_1(2{\rm P}) & = & \alpha^2 \, \frac{4}{27 m^4} 
\left(E_{\rm 2P} - E_{\rm 1S}\right)^2 \, 
  |\langle {\rm 1S} | \bbox{p} | {\rm 2P} \rangle|^2 \,
    \left< {\rm 1S} \left| p^i \,   
       \frac{1}{H + E_{\rm 2P} - 2\,E_{\rm 1S}} \, p^i 
          \right| {\rm 1S} \right>
\nonumber\\[3ex]
& = & \frac{2^5}{3^9} \, \alpha^2 \, (Z\alpha)^6 \, m \, {\cal M}_1\,,
\end{eqnarray}
where the summation convention is used and the
matrix element ${\cal M}_1$ reads
\begin{equation}
{\cal M}_1 = \frac{1}{m} \,     
  \left< {\rm 1S} \left| \, p^i \,
    \frac{1}{H + E_{\rm 2P} - 2\,E_{\rm 1S}} \, p^i \,
      \right| {\rm 1S} \right> = 0.880
\end{equation}
and 
\begin{equation}
\left| \left< 1{\rm S} \left| \frac{\bbox{p}}{m} \right| 2{\rm P} \right>
\right|^2 = \frac{2^9}{3^8} \, (Z\alpha)^2\,.
\end{equation}
Note that the contribution $C_1$ lacks the factors $\pi$ in the denominator
which are characteristic of other two--loop corrections: these are compensated
by additional factors of $\pi$ in the numerator that characterize the 
pole contributions.

The rainbow diagram in Fig.~\ref{fig1} (b) 
with the second loop inside the first 
does not create squared imaginary contributions.
From the irreducible part of the
loop-after-loop diagram in Fig.~\ref{fig1} (c) (we exclude the reference state
in the intermediate electron propagator), we obtain
\begin{eqnarray}
\label{T2}
\lefteqn{{\mathcal T}_2(2{\rm P})
= \lim_{\delta \to 0^+}
- \left( \frac{2 \, \alpha}{3 \,\pi\,m^2} \right)^2 \,
\int_0^{\epsilon_1} {\rm d}\omega_1 \, \omega_1 \,
\int_0^{\epsilon_2} {\rm d}\omega_2 \, \omega_2 \,} \nonumber\\[2ex]
& & \times
\left< 2{\rm P} \left|  p^i \, 
\frac{1}{H - {\rm i}\, \delta - E_{\rm 2P} + \omega_1} \, p^i \,
\left( \frac{1}{H - E_{\rm 2P}} \right)' \, p^j \,
\frac{1}{H - {\rm i}\, \delta - E_{\rm 2P} + \omega_2} \, p^j 
\right| 2{\rm P} \right>\,.
\end{eqnarray}
Again, picking up only those terms which are
generated by the infinitesimal half-circles around the poles
at $\omega_1 = E_{\rm 2P} - E_{\rm 1S}$ and
$\omega_2 = E_{\rm 2P} - E_{\rm 1S}$,
we obtain the contribution $C_2(2{\rm P})$ involving
squared decay rates,
\begin{eqnarray}
\label{C2}
C_2(2{\rm P}) & = & \alpha^2 \, \frac{4}{27 m^4}
\left(E_{\rm 2P} - E_{\rm 1S}\right)^2 \,
  |\langle {\rm 1S} | \bbox{p} | {\rm 2P} \rangle|^2 \,
    \left< {\rm 1S} \left| p^i \,
       \left( \frac{1}{H - E_{\rm 2P}} \right)' \, p^i
          \right| {\rm 1S} \right>
\nonumber\\[3ex]
& = & \frac{2^5}{3^9} \, \alpha^2 \, (Z\alpha)^6 \, m \, {\cal M}_2\,,
\end{eqnarray}
where the matrix element ${\cal M}_2$ reads
\begin{equation}
{\cal M}_2 = \frac{1}{m} \,
  \left< {\rm 1S} \left| \, p^i \,
    \left( \frac{1}{H - E_{\rm 2P}} \right)' \, p^i \,
      \right| {\rm 1S} \right>
  = \frac{6952}{6561} + \frac{4096}{2187} \, \ln\left(\frac{9}{8}\right) 
  = 1.28\,.
\end{equation}
The prime in the reduced Green function
indicates that the 2P state is excluded from the
sum over intermediate states. 

This contribution deserves a more detailed discussion because
$C_2$ can be written as a ``second-order perturbation'' according to
\begin{equation}
\label{2ndorder}
C_2 = \left< 2{\rm P} \left| 
\left\{ {\rm i} \,{\rm Im}\,
\Sigma^{\rm (1L)}_{\rm NR}(E_{\rm 2P}) \right\} \,
\left( \frac{1}{E_{\rm 2P} - H} \right)'  \, 
\left\{ {\rm i} \,{\rm Im}\, 
\Sigma^{\rm (1L)}_{\rm NR}(E_{\rm 2P}) \right\} \,
\right | 2{\rm P} \right>\,,
\end{equation}
where $\Sigma^{\rm (1L)}_{\rm NR}(E)$ denotes the one-loop 
nonrelativistic self-energy operator with the energy argument $E$;
the matrix elements of this operator read
\begin{equation}
\label{1loff}
\left< \phi_1 \left| \Sigma^{\rm (1L)}_{\rm NR}(E) \right| \phi_2 \right> =
\lim_{\delta \to 0^+} - \frac{2 \alpha}{3} \,
\int_0^\epsilon {\rm d}\omega \, \omega \,
\left< \phi_1 \left| \frac{\bbox{p}}{m} \,
\left( \frac{1}{H - {\rm i}\,\delta - E + \omega} \right) \,
\frac{\bbox{p}}{m} \right| \phi_2 \right>\,.
\end{equation}
In (\ref{2ndorder}) and (\ref{1loff}), 
we pick up only the term corresponding to the square of the two imaginary 
contributions and obtain the expression
\begin{eqnarray}
\sum_{n \neq 2} 
& & \left( - {\rm i}\, \frac{2 \alpha}{3} \,
\left( E_{\rm 2P} - E_{\rm 1S} \right) \,
\left< 2{\rm P} \left| \frac{p^i}{m} \right| 1{\rm S} \right> \,
\left< 1{\rm S} \left| \frac{p^i}{m} \right| n{\rm P} \right> \right) \,
\frac{1}{E_{\rm 2P} - E_{\rm nP}} \nonumber\\[2ex]
& & \quad \left( - {\rm i}\, \frac{2 \alpha}{3} \,
\left( E_{\rm 2P} - E_{\rm 1S} \right) \,
\left< n{\rm P} \left| \frac{p^j}{m} \right| 1{\rm S} \right> \,
\left< 1{\rm S} \left| \frac{p^j}{m} \right| 2{\rm P} \right> \right) \,,
\end{eqnarray}
where the sum covers all P states except the reference state
(both discrete and continuum states).
Now, we complete the sum over the intermediate states, observing that
only P states yield a nonvanishing contribution.
After angular averaging over atomic momenta, we obtain a contribution of
\begin{equation}
\frac{4}{27} \, \alpha^2 \,
\left( E_{\rm 2P} - E_{\rm 1S} \right)^2 \,
\left| \left< 1{\rm S} \left| \frac{\bbox{p}}{m} \right| 2{\rm P} \right> 
\right|^2 \,
\left< 1{\rm S} \left| \frac{p^j}{m} \,
\left( \frac{1}{H - E_{\rm 2P}} \right)'  \,
\frac{p^j}{m} \right | 1{\rm S} \right>
\end{equation}
in agreement with (\ref{C2}).

From the derivative term (reducible part of the loop-after-loop
diagram), we obtain
\begin{eqnarray}
\label{T3}
\lefteqn{{\mathcal T}_3(2{\rm P}) 
= \lim_{\delta \to 0^+}
\left( \frac{2 \, \alpha}{3 \,\pi\,m^2} \right)^2 \,
\int_0^{\epsilon_1} {\rm d}\omega_1 \, \omega_1 \,
\int_0^{\epsilon_2} {\rm d}\omega_2 \, \omega_2 \,} \nonumber\\[2ex]
& &
\times
\left< 2{\rm P} \left|  p^i \, 
\frac{1}{H - {\rm i}\,\delta - E_{\rm 2P} + \omega_1} \, p^i \,
\right| 2{\rm P} \right> \,
\left< 2{\rm P} \left|  p^j \, 
\left( \frac{1}{H - {\rm i}\,\delta - E_{\rm 2P} + \omega_2} \right)^2 \, 
p^j \right| 2{\rm P} \right>\,,\\[3ex]
\label{C3}
C_3(2{\rm P}) & = & - \alpha^2 \, \frac{4}{9 m^4}
\left(E_{\rm 2P} - E_{\rm 1S}\right) \,
  |\langle {\rm 1S} | \bbox{p} | {\rm 2P} \rangle|^4 \,
= -\frac{1}{4} \, \frac{\Gamma^2_{\rm 2P}}{E_{\rm 2P} - E_{\rm 1S}}
= - \frac{2^{17}}{3^{17}} \, \alpha^2 \, (Z\alpha)^6 \, m \,,
\end{eqnarray}
where $\Gamma_{\rm 2P} = (2/3)^8 \, \alpha \, (Z\alpha)^4 \, m$ 
is the decay width of the 2P state.
The last contribution of the ``squared-decay'' type
-- it originates from the ``seagull term'' characteristic of
NRQED -- reads
\begin{eqnarray}
\label{T4}
{\mathcal T}_4(2{\rm P})
&=& \lim_{\delta \to 0^+} \left( \frac{2 \, \alpha}{3 \,\pi\,m^2} \right)^2 \,
\int_0^{\epsilon_1} {\rm d}\omega_1 \, \omega_1 \,
\int_0^{\epsilon_2} {\rm d}\omega_2 \, \omega_2 \, \nonumber\\[2ex]
& &
\times
\left< 2{\rm P} \left|  p^i \, 
\frac{1}{H - {\rm i}\,\delta - E_{\rm 2P} + \omega_1} \,
\frac{1}{H - {\rm i}\,\delta - E_{\rm 2P} + \omega_2} \, 
p^i \right| 2{\rm P} \right>\,, \\[3ex]
\label{C4}
C_4(2{\rm P}) & = & - \alpha^2 \, \frac{4}{9 m^3}
\left(E_{\rm 2P} - E_{\rm 1S}\right)^2 \,
  |\langle {\rm 1S} | \bbox{p} | {\rm 2P} \rangle|^2 \,
= -\frac{3^9}{2^{14}} \, \frac{\Gamma^2_{\rm 2P}}{E_{\rm 2P} - E_{\rm 1S}}
\nonumber\\[3ex]
& = & - \frac{2^{5}}{3^{8}} \, \alpha^2 \, (Z\alpha)^6 \, m\,.
\end{eqnarray}
Adding all contributions, we obtain a shift of
\begin{equation}
\label{b60}
\sum_{i=1}^4 C_i(2{\rm P}) = \left(\frac{\alpha}{\pi}\right)^2 \,
\frac{(Z\alpha)^6 \, m}{2^3} \, (-0.188)\,.
\end{equation}
for the 2P level. For atomic hydrogen ($Z=1$), this correction
amounts to $-14.9~{\rm Hz}$. Compared to the total Lamb shift
of the 2P level, for which a value of $-12\,638\,380(80)~{\rm Hz}$
has been given in~\cite{JePa1996} (for the 
$2{\rm P}_{1/2}$-state), this is a tiny effect. 
The comparison illustrates the accuracy of the predictions obtained
using the Gell--Mann--Low--Sucher theorem~(\ref{GMLow}).
The theoretical value of the 2P Lamb shift has been improved in part by the 
recent numerical evaluation of the one-loop self energy for 
low $Z$~\cite{JeMoSo2001pra}, but the removal of the
main theoretical uncertainty will necessitate the complete evaluation
of the two-loop self-energy corrections of order 
$\alpha^2\,(Z\alpha)^6$. Currently, only the double logarithm is known
in this order in the $(Z\alpha)$-expansion~\cite{Ka1996,JeNa2002},
and the groundwork for the evaluation of single-logarithmic
and nonlogarithmic corrections has been laid in~\cite{Pa2001}.
In Eq.~(\ref{b60}), we write the final result
as a contribution to the $B_{60}$ coefficient
(for a definition of the analytic $B$-coefficients
of the two-loop self-energy, see e.g.~\cite{JePa2002}).

For the 3P state, we have to take into account the decays
into the 1S and 2S states. For example, the contribution $C_1(3{\rm P})$ reads
\begin{eqnarray}
C_1(3{\rm P}) & = & \alpha^2 \, \frac{4}{27 m^4} \, \left\{
\left(E_{\rm 3P} - E_{\rm 1S}\right)^2 \, 
  |\langle {\rm 1S} | \bbox{p} | {\rm 3P} \rangle|^2 \,
    \left< {\rm 1S} \left| p^i \,   
       \frac{1}{H + E_{\rm 3P} - 2\,E_{\rm 1S}} \, p^i 
          \right| {\rm 1S} \right> \right.
\nonumber\\[1ex]
& & \quad \left. 
+ \left(E_{\rm 3P} - E_{\rm 2S}\right)^2 \, 
  |\langle {\rm 2S} | \bbox{p} | {\rm 3P} \rangle|^2 \,
    \left< {\rm 2S} \left| p^i \,   
       \frac{1}{H + E_{\rm 3P} - 2\,E_{\rm 2S}} \, p^i 
          \right| {\rm 2S} \right> \right\}
\nonumber\\[1ex]
& & + \alpha^2 \, \frac{8}{27 m^4} \, \chi \,
{\rm Re}\left(\langle {\rm 3P} | p^j | {\rm 1S} \rangle \,
\left< {\rm 1S} \left| p^i \,   
  \frac{1}{H + E_{\rm 3P} - E_{\rm 2S} - E_{\rm 1S}} \, p^i 
     \right| {\rm 2S} \right> \,
\langle {\rm 2S} | p^j | {\rm 3P} \rangle
\right)
\nonumber\\[3ex]
& = & \left(\frac{\alpha}{\pi}\right)^2 \,
\frac{(Z\alpha)^6 \, m}{3^3} \, (0.135)\,,
\end{eqnarray}
where $\chi = (E_{\rm 3P} - E_{\rm 1S}) \, (E_{\rm 3P} - E_{\rm 2S})$.
The sum of $C_1$ -- $C_4$ for the 3P state of atomic hydrogen is
\begin{equation}
\label{res3Pstate}
\sum_{i=1}^4 C_i(3{\rm P}) = \left(\frac{\alpha}{\pi}\right)^2 \,
\frac{(Z\alpha)^6 \, m}{3^3} \, (-0.319)\,.
\end{equation}
For atomic hydrogen, this correction evaluates to $-7.47~{\rm Hz}$.

%
% Interpretation of the Squared Decay Rate
%
\section{Interpretation of the Squared Decay Rate}
\label{Interpretation}

In this section, we will investigate the question of whether
the squared decay rates receive a natural interpretation 
within the formalism of scattering theory. 

\begin{figure}[htb!]
\begin{center}
\begin{minipage}{14cm}
\centerline{\mbox{\epsfysize=15.0cm\epsffile{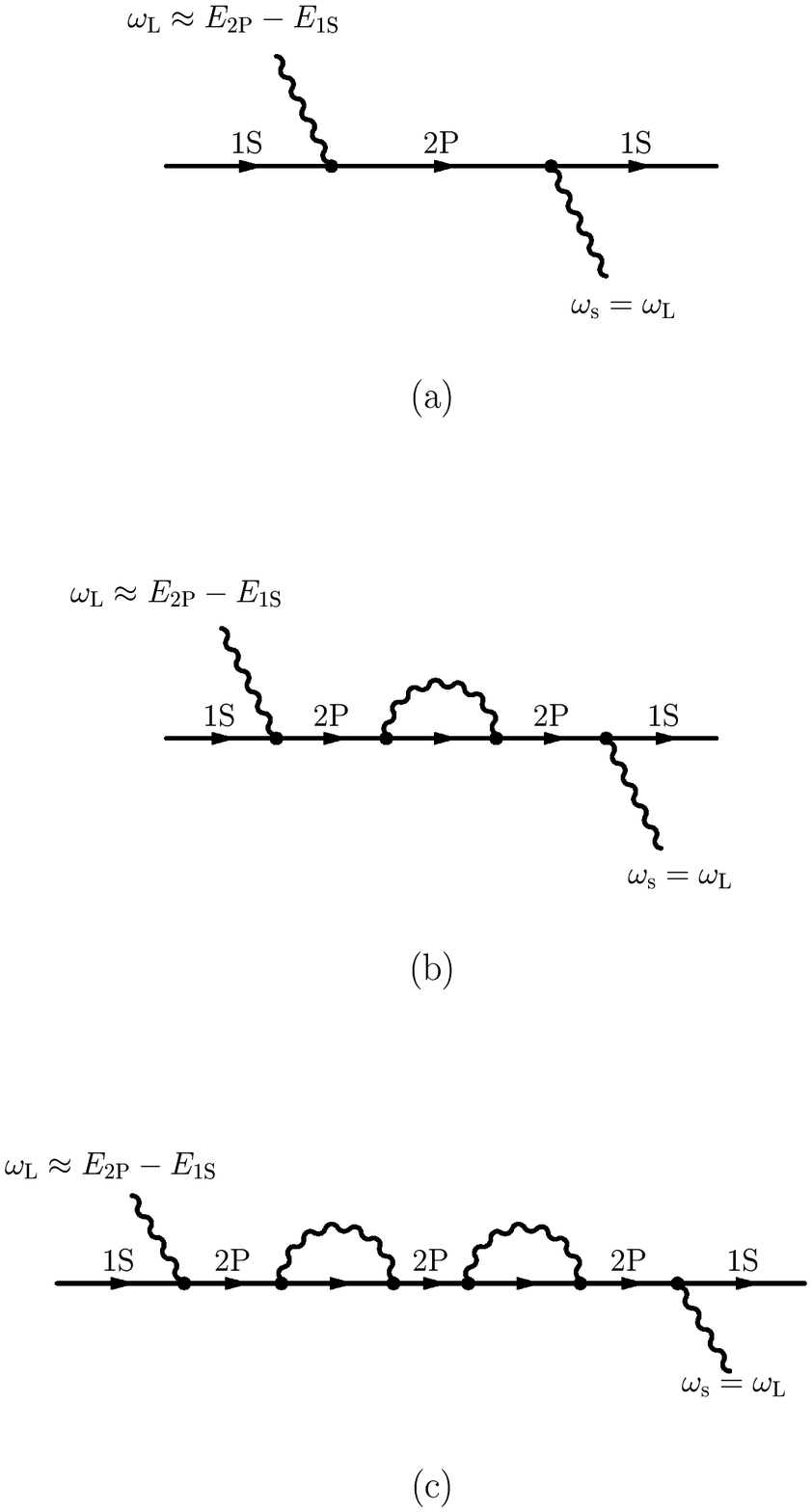}}}
\caption{\label{fig2} Contributions to the photon scattering
cross section near the 2P resonance of atomic hydrogen.
The laser photon of frequency $\omega_{\rm L} \approx E_{\rm 2P} - E_{\rm 1S}$
is absorbed, and the dominant contribution to the scattering amplitude
is from the virtual 2P state, as indicated by the label on the 
electron line. The spontaneously emitted photon has a 
frequency $\omega_{\rm s}$. Figs.~(b) and (c) represent radiative corrections
to the scattering amplitude. The arrow of time is from left to right.}
\end{minipage}
\end{center}
\end{figure}

We consider the scattering amplitude associated with the diagram
in Fig.~\ref{fig2}. The hydrogenic atom in the ground state is 
excited by a laser photon with frequency $\omega_{\rm L}$ that 
is close to the resonance $\omega_{\rm L} \approx E_{\rm 2P} -
E_{\rm 1S}$. Within the resonance approximation, we may restrict
the sum over intermediate states to the 2P level only, as indicated
by the label ``2P'' for the electron line. This is the so-called
resonance approximation.

%
% Fig. 2
%

Fig.~\ref{fig2} (a) represents the dominant 
Kramers--Heisenberg contribution~\cite{KrHe1925} without radiative 
corrections. In Fig.~\ref{fig2} (b), we have a one-loop self-energy
insertion in the electron propagator. We assume that the 
absorbed photon is very close to the resonance 
($\omega_{\rm L} \approx E_{\rm 2P} - E_{\rm 1S}$) and 
set the energy argument of the self-energy insertion equal
to $E_{\rm 2P}$. Using these approximations,
the self-energy insertion, within the
resonance approximation, can be described by the diagonal matrix
element 
$\left< 2{\rm P} \left| 
\Sigma^{\rm (1L)}_{\rm NR}(E_{\rm 2P}) \right| 2{\rm P} \right>$
of the self-energy operator (\ref{1loff}). The imaginary
part of this matrix element reads
\begin{equation}
{\rm i}\,{\rm Im} 
\left< 2{\rm P} \left| 
\Sigma^{\rm (1L)}_{\rm NR}(E_{\rm 2P}) \right| 2{\rm P} \right>
= 
- {\rm i}\, \frac{\Gamma_{\rm 2P}}{2}\,,
\end{equation}
where 
\begin{equation}
\Gamma_{\rm 2P} = \frac{2^8}{3^8}\alpha(Z\alpha)^4\,m
\end{equation}
is the well-known decay rate of the 2P state. The diagram in 
Fig.~\ref{fig2} (c) involves two one-loop self-energy insertions and
entails in that sense the ``square of the decay rate'' of the 2P level
within the resonance approximation. This square of the decay is usually
not interpreted as an energy shift. Rather, one sums the infinite series
of one-loop insertions of which the first terms
are shown in the Feynman diagrams in Figs.~\ref{fig2} (a,b,c).
We ignore in the sequel the real part of the self-energy operator
and define the ``decay rate operator'' $\hat\Gamma$ via the relation
\begin{equation}
\label{hatGamma}
{\rm i}\,{\rm Im}\,
\left< \phi_1 \left| \Sigma^{\rm (1L)}_{\rm NR}(E) \right| \phi_2 \right> =
-\frac{\rm i}{2}\,
\left< \phi_1 \left| \hat\Gamma(E) \right| \phi_2 \right>
\end{equation}
so that
\begin{equation}
-\frac{\rm i}{2}\,
\left< 2{\rm P} \left| \hat\Gamma(E_{\rm 2P}) \right| 2{\rm P} \right> =
- {\rm i}\, \frac{\Gamma_{\rm 2P}}{2}\,.
\end{equation}
This leads to the following infinite series representing the 
electron propagator in the resonance approximation with an 
infinite number of one-loop self-energy insertions (of which 
we discard the real part and keep only the imaginary part
$- {\rm i}\, \Gamma_{\rm 2P}/2$),
\begin{eqnarray}
\label{InfSer}
\lefteqn{
\frac{| 2{\rm P} \rangle \, \langle 2{\rm P} |}
  {E_{\rm 2P} - (E_{1S} + \omega_{\rm L})}
- \frac{| 2{\rm P} \rangle}
  {E_{\rm 2P} - (E_{1S} + \omega_{\rm L})} \, 
\left( - {\rm i}\, \frac{\Gamma_{\rm 2P}}{2} \right) \,
\frac{ \langle 2{\rm P} |}
  {E_{\rm 2P} - (E_{1S} + \omega_{\rm L})} }\nonumber\\[2ex]
& & + \frac{| 2{\rm P} \rangle}
  {E_{\rm 2P} - (E_{1S} + \omega_{\rm L})} \,
\left( - {\rm i}\, \frac{\Gamma_{\rm 2P}}{2} \right) \,
\frac{ | 2{\rm P} \rangle \, \langle 2{\rm P} | }
  {E_{\rm 2P} - (E_{1S} + \omega_{\rm L})} 
\left( - {\rm i}\, \frac{\Gamma_{\rm 2P}}{2} \right) \,
\frac{ \langle 2{\rm P} |}
  {E_{\rm 2P} - (E_{1S} + \omega_{\rm L})} + \dots \nonumber\\[2ex]
& & =  \frac{| 2{\rm P} \rangle \, \langle 2{\rm P} |}
  {E_{\rm 2P} - {\rm i}\, \Gamma_{\rm 2P}/2 - (E_{1S} + \omega_{\rm L})}\,.
\end{eqnarray}
This result expresses the well-known fact that the decay rate term
$- {\rm i}\, \Gamma_{\rm 2P}/2$ in the electron propagator denominator 
is generated by summing an infinite series of Feynman diagrams
involving one-loop self-energy insertions; the first three terms
in this series are shown in Figs.~\ref{fig2} (a) -- (c).

\begin{figure}[htb!]
\begin{center}
\begin{minipage}{14cm}
\centerline{\mbox{\epsfysize=11.0cm\epsffile{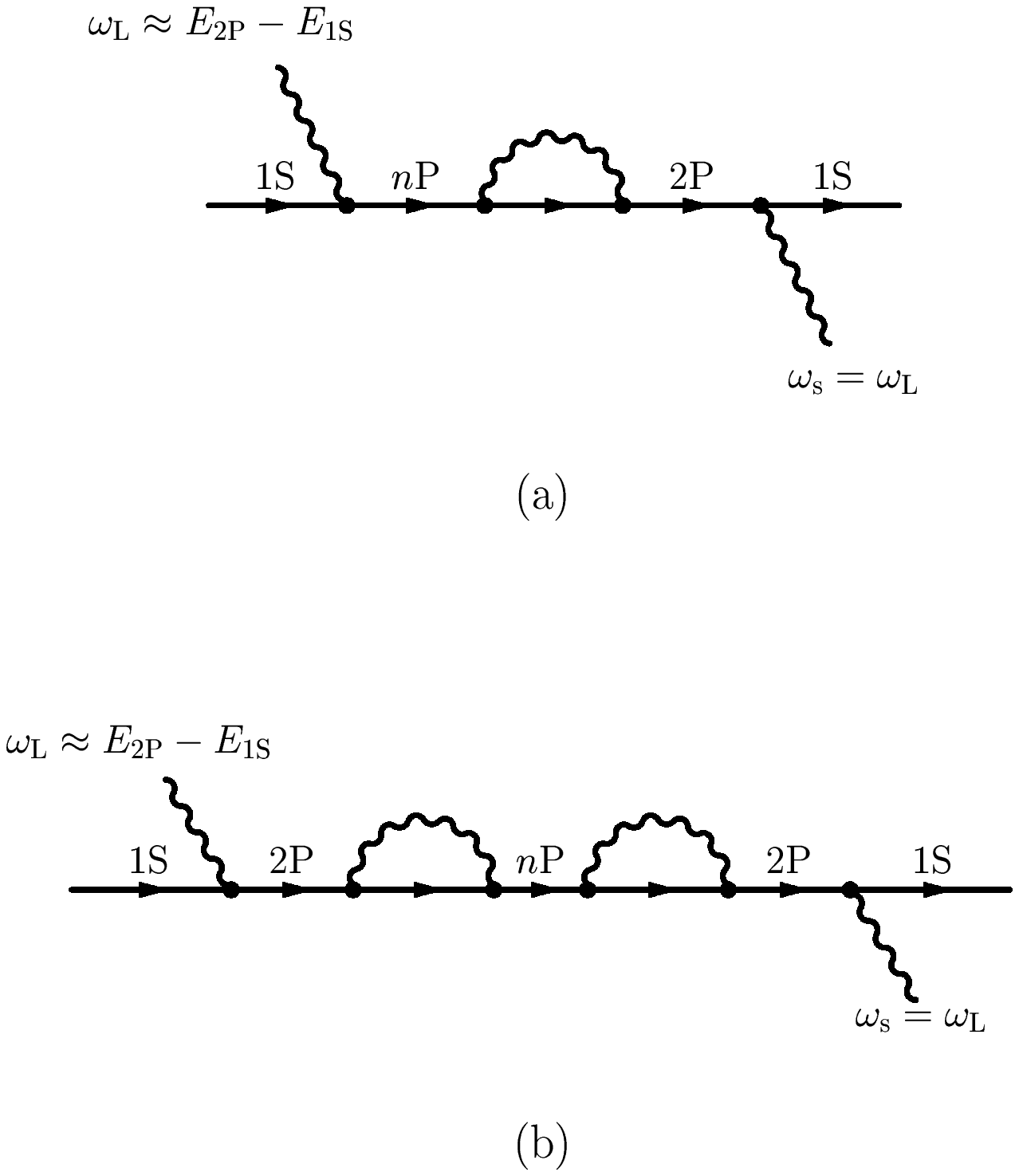}}}
\caption{\label{fig3} Same as Fig.~2, but with off-resonant
virtual states ($n$P states with $n \neq 2$). 
The arrow of time is from left to right.}
\end{minipage}
\end{center}
\end{figure}

Let us now go beyond the resonance approximation inherent 
to Eq.~(\ref{InfSer}) and consider off-resonant atomic
states (see Fig.~\ref{fig3}). We therefore replace
\begin{equation}
\label{repl}
\frac{| 2{\rm P} \rangle \, \langle 2{\rm P} |}
  {E_{\rm 2P} - {\rm i}\, \Gamma_{\rm 2P}/2 - (E_{1S} + \omega_{\rm L})}
\to
\frac{1}{H - {\rm i}\, {\hat\Gamma}/2 - (E_{1S} + \omega_{\rm L})}\,.
\end{equation}
That means that the resonance term involving only the 2P state
is replaced by the full Green function of the electron,
including the off-resonant states and the ``decay rate operator''
$\hat\Gamma$ defined in Eq.~(\ref{hatGamma}). The ``energy argument''
of $\hat\Gamma \equiv \hat\Gamma(E)$ is to be taken as 
$E = E_{1S} + \omega_{\rm L} \approx E_{\rm 2P}$. 
In Eq.~(\ref{repl}), $H$ is {\em a priori} the Schr\"{o}dinger
Hamiltonian for the bound electron [we have done all calculations
in the nonrelativistic (NR) approximation, but the calculations
may be generalized to the relativistic case]. 

%
% Fig. 3
%

It is now easy to verify by
inspection that the amplitudes corresponding to the Feynman 
diagrams in Figs.~\ref{fig3} (a) and (b) are generated 
by an expansion of the propagator 
$1/\{H - {\rm i}\, {\hat\Gamma}/2 - (E_{1S} + \omega_{\rm L})\}$
in powers of ${\hat\Gamma}$. Furthermore, the diagram
in Fig.~\ref{fig3} (b) exactly corresponds to the 
``second-order perturbation'' in Eq.~(\ref{2ndorder}).
So, we conclude that the energy shifts by squared decay rates
should rather be interpreted as radiative nonresonant
corrections to the photon scattering cross section than energy shifts
of individual atomic levels. Specifically, the ``second-order perturbation'' 
in Eq.~(\ref{2ndorder}), which corresponds to the diagram in
Fig.~\ref{fig3} (b), is merely a nonresonant generalization of the 
resonant diagram Fig.~\ref{fig2} (c). Both diagrams --
Fig.~\ref{fig2} (c) and Fig.~\ref{fig3} (b) -- can be treated in a 
natural way by appropriate replacements within the electron
propagator denominator. Specifically, we have the replacement
$H \to H - {\rm i}\,\hat\Gamma/2$ for the infinite series involving
the diagrams in Figs.~\ref{fig3} (a) and (b), and the 
corresponding replacement $E_{\rm 2P} \to E_{\rm 2P} - 
{\rm i}\,\Gamma_{\rm 2P}/2$ for the propagator denominator in the 
resonance approximation [see Figs.~\ref{fig2} (b) and (c) and 
the right-hand side of Eq.~(\ref{InfSer})].
Both added terms ($- {\rm i}\,\hat\Gamma/2$ and
$- {\rm i}\,\Gamma_{\rm 2P}/2$) are {\em not} real energy shifts.

%
% Conclusions
%
\section{Conclusions}
\label{Conclusions}

We have investigated a problematic set of two-loop
self-energy corrections involving the square of the decay rate
of the atomic state. 
These self-energy corrections cannot be interpreted as 
radiative energy shifts in the usual sense, although the relevant terms
are generated by a fourth-order expansion of the 
Gell--Mann--Low--Sucher theorem (\ref{GMLow}) in powers of the
quantum electrodynamic interaction Lagrangian (\ref{Hint}).
As explained in Sec.~\ref{2Pstate}, some of 
the expressions arise naturally if we treat the
self-energy leading to the decay rate as a first-order perturbation and
consider the resulting second-order perturbation.
The same problematic expressions involving squared decay rates result 
from the alternative formalism for deriving level shifts based on the
two-time Green function method~\cite{Sh2002}. 
At some risk to over-simplification,
we can state that the difficulties are related
to the fact that the current methods inadequately address the
question of the preparation of the excited atomic states
and involve asymptotic states with an infinite lifetime.
As explained in Sec.~\ref{Interpretation}, some of the problematic 
two-loop self-energy corrections find a
natural interpretation as radiative nonresonant corrections
to the line-shape for atomic transitions which are
of second order in the ``decay rate operator'' $\hat{\Gamma}$
defined in Eq.~(\ref{hatGamma}).

For the 2P--level in atomic hydrogen, as shown in
Sec.~\ref{2Pstate}, the problematic
energy shift by squared
decay rates is $-14.9~{\rm Hz}$ [see Eq.~(\ref{b60})].
For the 3P state of atomic hydrogen, the correction amounts to
$-7.47~{\rm Hz}$ [see Eq.~(\ref{res3Pstate})].
It is perhaps interesting to note that the effect discussed
here scales as $\alpha^2 \, (Z\alpha)^6 \, m \, c^2$
(for P states and states with higher angular momenta).
The same order-of-magnitude is characteristic of the shift of the 
peak of the total photon scattering cross section in electric-dipole
transitions in atomic hydrogen~\cite{Lo1952,LaSoPlSo2001,JeMo2001,LaSoPlSo2002}.
As is evident from Eqs.~(\ref{C3}) and (\ref{C4}),
an order-of-magnitude estimate for the problematic
two-loop effect is given
by the ratio of the square of the decay rate of the atomic state
to a typical atomic energy level difference.
For the 2S state with a natural line width of $1.3~{\rm Hz}$,
the squared decay rate therefore does not represent an appreciable 
predictive limit in current and future experiments. The situation is different
for the 2S--8D transition studied in~\cite{BeEtAl1997}. Parametrically,
the ``squared-decay'' corrections are of the order of
$\alpha^2 \, (Z\alpha)^6$ for the 8D state; however, if we
assume a typical $1/n^3$-type scaling of the effect ($n$
is the principal quantum number), then we immediately
obtain an estimate below $1~{\rm Hz}$ for the problematic
effect in the 2S--8D transition in atomic hydrogen.

There has recently been a dramatic increase in the
accuracy to which atomic energy levels can be measured
experimentally~\cite{NiEtAl2000} and evaluated 
theoretically (e.g.~\cite{Pa1994prl,JeMoSo1999,MeRi2000,Ye2001,Pa2001}).
Tiny nonresonant effects that influence the
natural line shape of hydrogenic transitions have received considerable
attention~\cite{LaSoPlSo2001,JeMo2001,LaSoPlSo2002}, and it
has been
pointed out that nonresonant effects are enhanced in differential cross
sections as opposed to total cross sections~\cite{JeMo2001,LaSoPlSo2002}.
Essentially, the nonresonant effects give rise to a process-dependent
new line shape, different from the Lorentzian, which has to be fitted by
a number of parameters. By a suitable fit of the line shape,
taking into account properly the relevant
experimental conditions, it is {\em in principle}
possible to correct the observed peak of the cross section for the
nonresonant contributions. The same applies at least in part 
to the problematic two-loop corrections discussed here (see 
the discussion in Sec.~\ref{Interpretation}), provided
they are interpreted properly as radiative corrections to the 
off-resonance effects on the line shape. 
The problematic two-loop corrections illustrate
that it is impossible to separate 
atomic energy levels at the order of $\alpha^2 \, (Z\alpha)^6 \, m\,c^2$,
that is to say to define an energy shift that relates to 
one and only one level.

In~\cite{Pa1991}, it has been stressed that
several important problems associated with the normalization
of electronic states and with the infrared catastrophe can be avoided 
if we consider the electron in a free state as obtained by 
ionization of a bound state (for example, by ionization of
the atomic {\em ground} state
which does not decay at all and is therefore
the only ``true'' asymptotic state). 
The considerations presented in~\cite{Pa1991} imply
that infrared divergences of free-electron quantum electrodynamics
originate because one ``disregards the methods of obtaining
and detecting these (free) states''.
It has been shown that in special cases, 
the spontaneous emission from a certain
``well'' constructed state is exactly exponential~\cite{Pa1991}.
We have mentioned above
that the interpretation of the energy shift by squared decay rates 
is problematic because it is calculated as a matrix element 
evaluated on the excited atomic state. Within the 
formalism introduced in~\cite{Pa1991}, we can argue
that this procedure of evaluating the energy shift of the
excited atomic state according to the current formalism is
inconsistent because it disregards the fact that according 
to~\cite{Pa1991}, the only
``proper'' way of defining the excited state is to view this state
as having been obtained by excitation from the ground state.

%
% Acknowledgements
%
\section*{Acknowledgements}

The authors acknowlegde insightful discussions with 
Professor Jean Zinn--Justin. U.D.J., J.E. and C.H.K.
gratefully acknowledge
funding by the Deutsche Forschungsgemeinschaft (Nachwuchsgruppe
within the Sonderforschungsbereich 276), and
K.P. acknowledges support from the
Polish Committee for Scientific Research under
contract no. 2P03B 057 18.

\end{document}